\documentclass{ws-ijmpb}

\begin{document}

\markboth{Mohammad M. Valizadeh and Sashi Satpathy}
{RKKY Interaction for the Spin Polarized Electron Gas}

%
\catchline{}{}{}{}{}
%

\title{RKKY Interaction for the Spin Polarized Electron Gas}

\author{Mohammad M. Valizadeh}

\address{Department of Physics $\&$ Astronomy, University of Missouri, Columbia, MO 65211, USA
\\
mvbr5@mail.missouri.edu
}

\author{Sashi Satpathy}

\address{Department of Physics $\&$ Astronomy, University of Missouri, Columbia, MO 65211, USA
\\
satpathys@missouri.edu}

\maketitle

\begin{history}
\end{history}

\begin{abstract}
We extend the original work of  Ruderman, Kittel, Kasuya, and Yosida (RKKY) on the interaction between two magnetic moments embedded in an electron gas to the case where the electron gas is spin polarized.  
The broken symmetry of a host material 
introduces the Dzyaloshinsky-Moriya (DM) vector 
and tensor interaction terms, in addition to the standard RKKY term,
 so that the net interaction energy has the form:   
$ {\cal H}  = J \vec S_1 \cdot \vec S_2 + \vec D \cdot   \vec S_1 \times \vec S_2 
+  \vec S_1 \cdot  \stackrel{\leftrightarrow}{\Gamma} \cdot \vec S_2    $.
We find that for the spin-polarized electron gas, a non-zero tensor interaction $\stackrel{\leftrightarrow}{\Gamma}$ is present in addition to the scalar RKKY interaction $J$, while $\vec D$ is zero due to the presence of inversion symmetry. 
Explicit    expressions for these are derived for the electron gas both in 2D and 3D. The RKKY interaction exhibits a beating pattern, caused by the presence of the two Fermi momenta $k_{F\uparrow}$ and $k_{F\downarrow}$,
while the $R^{-3}$ distance dependence of the original RKKY result for the 3D electron gas is retained. 
This model serves as a simple example of the magnetic interaction in systems with broken symmetry, which goes beyond the RKKY interaction.

\end{abstract}


\section{Introduction}

The Ruderman-Kittel-Kasuya-Yosida (RKKY) interaction between two magnetic moments embedded in a solid 
was originally derived for the free electron gas,
\cite{RKKY1,RKKY2,RKKY3} and it has been extensively studied since then.
 It has the Heisenberg form $ J  \vec S_1 \cdot \vec S_2$ and   shows an oscillatory behavior as a function of the distance $R$ between the two magnetic moments; 
 for example, $J(R) \propto  (x \cos x - \sin x ) / x^4$ for the electron gas in three dimensions (3D), where $ x = 2 k_F R$ with $k_F$ being the Fermi momentum.
Using lattice models, Dzyaloshinsky and Moriya (DM) showed that\cite{DM,Moriya} in situations with broken  symmetry (inversion or time reversal), there appears a vector or a tensor interaction term, $\vec D$ and $ \stackrel{\leftrightarrow}{\Gamma}$, respectively, in addition to the scalar, RKKY interaction $J$, so that the net interaction energy  is given by the Hamiltonian
$ {\cal H}  = J \vec S_1 \cdot \vec S_2 + \vec D \cdot   \vec S_1 \times \vec S_2 
+  \vec S_1 \cdot  \stackrel{\leftrightarrow}{\Gamma} \cdot \vec S_2    $.
There is currently a considerable interest on this less-studied DM interactions, because of the recent observation\cite{Tokura,sk1,sk2} of skyrmions, a novel vortex-like spin structure\cite{skyrmea1962}, that forms due to the competition between the RKKY and the DM terms.

While the RKKY interaction is quite simply illustrated by taking the example of the free electron gas, this is not the case for the DM interactions. The original work of DM involved lattice models, while in many subsequent works, complicated interactions such as
non-magnetic spin-flip centers\cite{Fert}   or relativistic scattering potentials\cite{Gyorffy} were used to obtain the DM interactions. 
In this paper, we point out that the simple generalization of the electron gas to include spin polarization does indeed produce the DM interaction,
{\it albeit} just the tensor part as the vector part $\vec D =0 $  because the inversion symmetry is still present.
%

\begin{figure}
\centering
\includegraphics[angle=0,width=0.55 \linewidth]{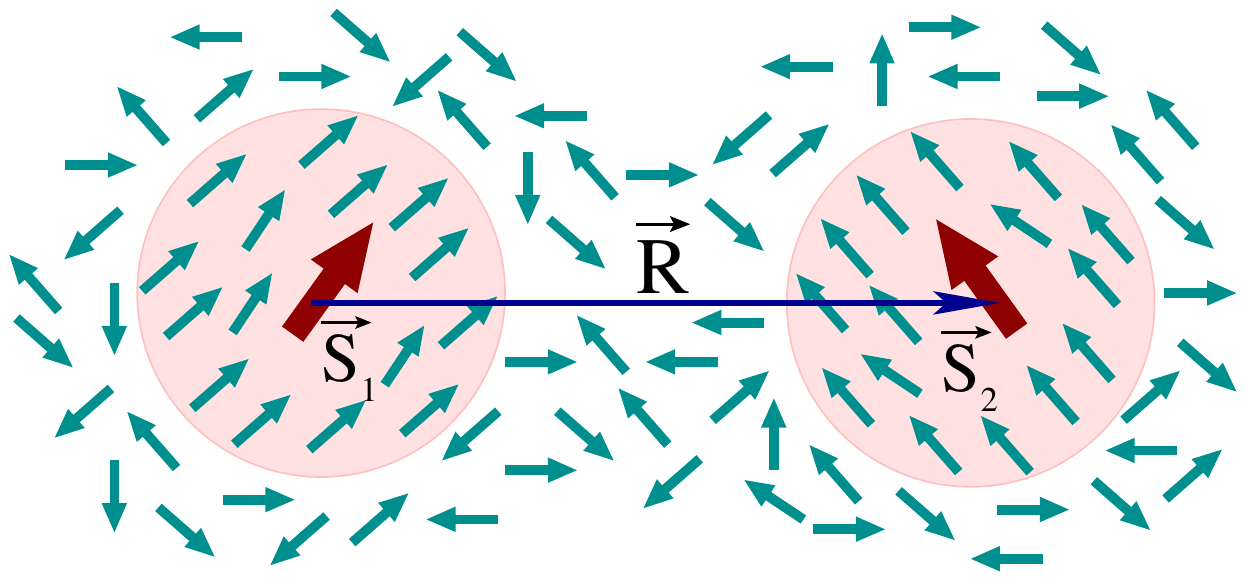} 
\caption{(Color online) 
The interaction between two magnetic moments mediated by the host electrons in the crystal, which leads to the well known RKKY interaction of the type $J \vec S_1 \cdot \vec S_2$. However,
if  the host crystal has broken symmetry, then additional  interaction terms appear, known as the Dzyaloshinsky-Moriya interaction. 
}
\label{RKKY}
\end{figure}

Before we discuss the case of the spin-polarized electron gas, we note that 
the coupling between the spin and orbital degrees of freedom is essential to the DM interactions, as without it electron energies are spin degenerate, $\varepsilon (k \uparrow) = \varepsilon (k \downarrow)$, and it turns out that, if this is true, then there would only be the RKKY interaction. This is the case for the standard, non-spin-polarized electron gas. 
If there is sufficient symmetry, even in the presence of spin-orbit coupling, the electron states may still be spin degenerate, so that the presence of the spin-orbit coupling alone is not enough  to produce the DM interaction.

In general, the presence of the inversion symmetry in a solid leads to the condition 
\begin{equation}
\varepsilon_{\vec k \uparrow} = \varepsilon_{- \vec k \uparrow},
\end{equation} 
while the time reversal symmetry
implies that
\begin{equation}
\varepsilon_{\vec k \uparrow} = \varepsilon_{- \vec k \downarrow},
\end{equation}
so that if both symmetries are present, one obtains the spin-degenerate energies, viz., $\varepsilon_{\vec k \uparrow} = \varepsilon_{  \vec k \downarrow}$. 
This equality is violated if there is at least one broken symmetry. 
The simplest way to include such a broken symmetry is to consider spin polarized bands in a 3D electron gas, which has broken time reversal symmetry. This may for example be produced by a magnetic field, either externally applied or produced by the constituent atoms in the solid. 
We now consider the interaction between magnetic moments placed in the spin-polarized electron gas.

\section{Spin-polarized electron gas and magnetic interaction}

The magnetic interaction energy between the two localized moments is obtained by using the quantum mechanical perturbation theory. The first magnetic moment perturbs the electron gas, which is then felt by the second moment, thereby coupling the two moments. 
The unperturbed eigenstates of the electron gas are plane wave states
\begin{align}
 & | \vec k \sigma    \rangle = \frac{1}{\sqrt \Omega} e^{i \vec k \cdot \vec r} |\sigma \rangle,     \nonumber \\
 &\varepsilon_{\vec k \sigma}    =  \frac{\hbar^2 k^2}{2m} \mp \Delta ,
\label{free-psi}	
\end{align}
where we have taken the energies to be spin dependent with
the $\mp$ sign corresponding to the energy of the spin $\uparrow$ and $\downarrow$
states, respectively,   and the box normalization has been used with $\Omega$ being the volume of the box. 
We now introduce two localized spins $\vec S_1$ and $\vec S_2$, located at the origin and at the position $\vec R$, respectively, which
interact with the electron gas via the contact interaction

\begin{align}
& V_1(\vec r) = - \lambda   \delta (\vec r ) \vec S_1 \cdot \vec s,        \nonumber \\
& V_2(\vec r) = - \lambda   \delta (\vec r - \vec R ) \vec S_2 \cdot \vec s,
\label{potential}
\end{align}
 where $\vec s$ is the spin of the electrons, and, in operator form, the perturbing potentials may be written as  
$\hat V_1=-\lambda| \vec r\rangle\langle\vec r| \vec S_1\cdot\vec s$ and similarly for $\hat V_2$.
%
\begin{figure}[h!]
\centering
\includegraphics[width=2.2in]{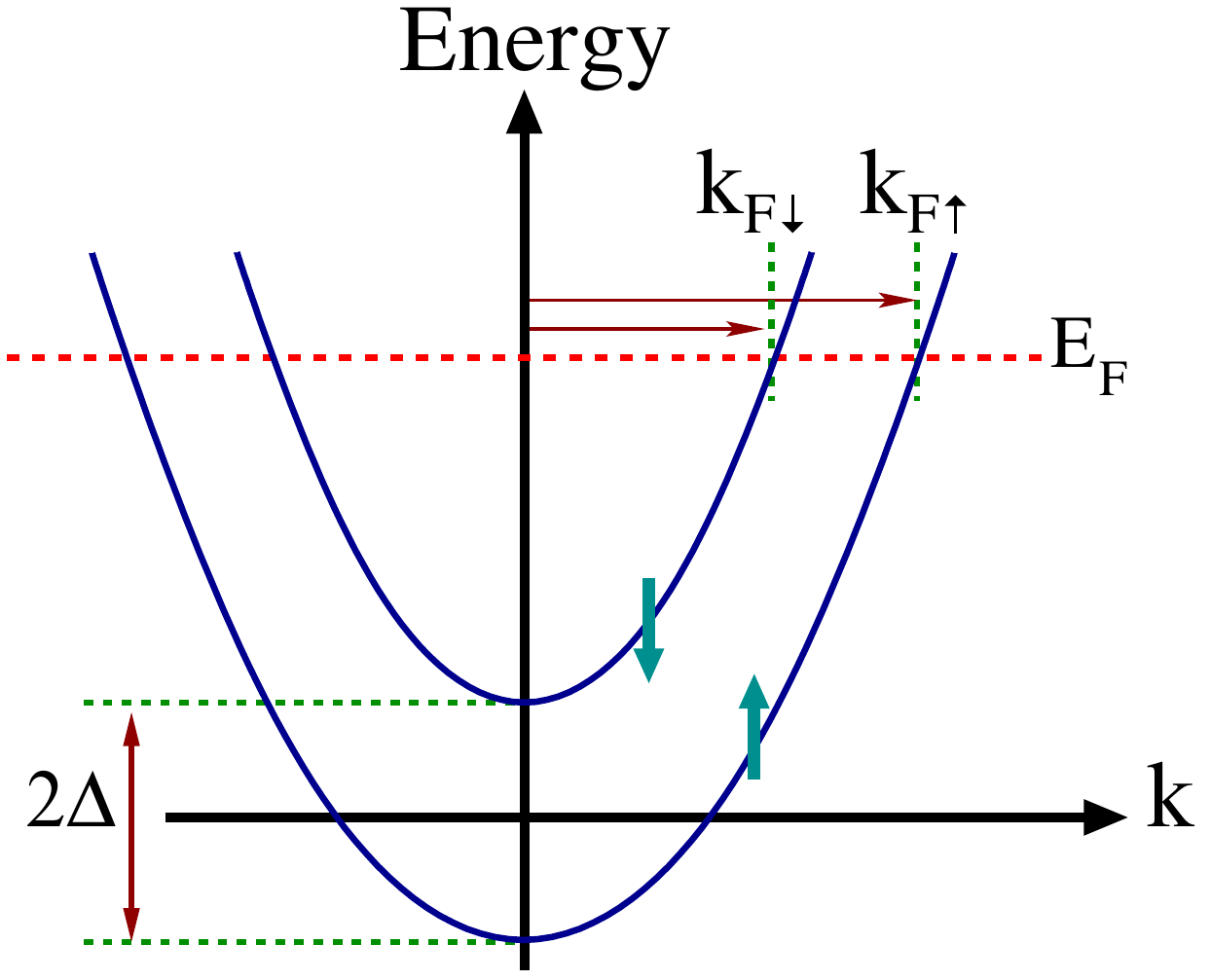}
\caption{ (Color online)  Electron gas with spin dependent energy. $E_F$ denotes the Fermi energy and the Fermi momenta in the two spin channels are indicated by $k_{F\uparrow}$
and $k_{F\downarrow}$.
}
\label{bands}
\end{figure}

The perturbed electron states $| \vec k \sigma \rangle \rangle$ due to  the first moment may be expressed in terms of the Lippmann-Schwinger equation\cite{Lippmann-Schwinger},
correct to the first order in perturbation theory
\begin{equation}
|\vec k\sigma\rangle\rangle=| \vec k\sigma\rangle+\hat G\hat V_1| \vec k\sigma\rangle,
\label{pert}
\end{equation}
where  $\hat G (E) = (E+ i \mu - \hat H)^{-1}$ with $\mu \rightarrow 0^+$  is the unperturbed retarded Green's function to be evaluated at the energy $E = \varepsilon_{\vec k \sigma}$, and $\hat H$ is the unperturbed Hamiltonian that produces the eigenstates in Eq. (\ref{free-psi}). Eq. (\ref{pert}) may be written in terms of the matrix elements of the Green's function, which reads
\begin{equation}
| \vec k\sigma  \rangle\rangle=   |  \vec k\sigma\rangle      -    \frac{ \lambda} {\sqrt{\Omega}}  
\int d^3r\sum    \limits_{\sigma^\prime}| \vec r\sigma^\prime    \rangle     G_{\sigma\sigma^\prime}        (\vec r,0,\varepsilon_{\vec k\sigma})
 {\vec S_{1}\cdot\vec s}_{\sigma\sigma^\prime}.
\label{pert2}
\end{equation}
%
%
The Green's function in the present case is diagonal in the spin indices 
\begin{align}
& G_{\sigma\sigma^\prime}(\vec r,\vec r \  ^\prime,E)   
\equiv   \langle\vec r \sigma|  \hat G(E)| \vec r \  ^\prime\sigma^\prime \rangle       \nonumber \\
& =\frac{1}{\Omega}   
\sum\limits_{\vec k}
\frac{e^{i\vec k\cdot(\vec r-\vec r \  ^\prime)}}   {E+i\mu -\varepsilon_{\vec k ,\sigma}}\delta_{\sigma\sigma^\prime}.
\label{Grr}
\end{align}
The spin $\vec S_2$ at $\vec R$ feels the perturbation via the contact interaction  $V_2 (r)$, leading to the 
interaction energy
\begin{equation}
E(\vec R)=\sum\limits_{\vec k \sigma}^{occ} \langle\langle\vec k\sigma| \hat V_2| \vec k\sigma\rangle\rangle=E_0+E(R)+O(\lambda^3).
\label{ER}
\end{equation}
This expression is evaluated by substituting Eq. (\ref{pert2}) for the perturbed states $| \vec k\sigma  \rangle\rangle$ and performing the summation over the occupied states.  
The $E_0$ term, which is $O(\lambda)$, introduces a constant shift in energy;  In fact there is an additional $O(\lambda)$
term, viz., $\langle\vec k\sigma| \hat V_1| \vec k\sigma\rangle$, originating from the perturbation of the first spin to the electron gas. 
Adding this to Eq. (\ref{ER}) to redefine $E_0$,  and performing a straightforward momentum summation, it is easy to show that, for the 3D electron gas,
\begin{equation}
E_0=\frac{-\hbar\lambda}{12\pi^2}(S_{1z}+ S_{2z})(k_{F\uparrow}^3-k_{F\downarrow}^3).
\label{E0-3D}
\end{equation}
A similar expression for the 2D electron gas is given later.

This result is easily interpreted, viz., that $E_0$ is the product of the energy gained per electron $\varepsilon_0$ times the excess number $N$ of
spin up electrons over  spin down electrons, where $\varepsilon_0 = -( \lambda  \hbar / 2\Omega) \times (S_{1z} + S_{2z})$, as computed for a single plane-wave electron
using the first-order perturbation theory with the perturbation Eq. (\ref{potential}),
and $N = (4\pi / 3) (k_{F\uparrow}^3-k_{F\downarrow}^3)/ (8\pi^3/\Omega)$. 
Note that we have used the standard Pauli matrices for the itinerant electron spins, e.g.,  
$s_z =  (\hbar  /2)   
 \begin{pmatrix} 
1 & 0 \\
0 & -1
 \end{pmatrix}$, etc.,
 so that effectively we have chosen $\hat z$ as the spin quantization axis and this is why the $z$ components $S_{1z}$ and $S_{2z}$ appear in Eq. (\ref{E0-3D}).
%

The next term in Eq. (\ref{ER}), which is $O(\lambda^2)$, produces the magnetic interaction and may be written as
\begin{equation}
E(R)=\frac{\lambda^2}{ \Omega}  \sum \limits_{\vec k \sigma}^{occ} \langle \sigma  |  \vec S_2\cdot\vec s  \ 
 G(\vec R, 0, \varepsilon_{\vec k \sigma})  \ 
\vec S_1\cdot\vec s \ | \sigma\rangle e^{-i\vec k\cdot\vec R}+ c.c.
\label{ER2}
\end{equation} 
Now, as mentioned already, $G_{\sigma^\prime \sigma^{\prime\prime}}   (\vec R, 0, \varepsilon_{\vec k \sigma})$ is diagonal in the spin space, though it can be different for  $\sigma = \uparrow$ or $\downarrow$ for a given $\vec k$.
It is convenient to write them in terms of the Pauli matrices:
%
%
\begin{align}     
G (\vec R, 0, \varepsilon_{\vec k \uparrow })  = 
g_0\sigma_0+ \frac{1}{2}  u(\sigma_0-\sigma_{z}), \nonumber \\
G (\vec R, 0, \varepsilon_{\vec k \downarrow })   =    g_0\sigma_0+\frac{1}{2}d(\sigma_0+\sigma_{z}),
\label{cases}
\end{align}
where $g_0$, $u$, and $d$ are c-numbers and functions of $\vec k$ and $\vec R$.
Putting this into  the energy expression Eq. (\ref{ER2}), we now get
\begin{align}
E  &  (R)=\frac{\lambda^2}{ \Omega} \left(  \sum\limits_{\vec k \sigma}^{occ}
 \langle\sigma| \vec S_2\cdot\vec s \  g_0\sigma_0 \  \vec S_1\cdot\vec s \  | \sigma\rangle e^{-i\vec k\cdot\vec R}       \right.   \nonumber\\
+&    \sum \limits_{k=0}^{k_{F\uparrow}}
 \langle\uparrow| \vec S_2\cdot\vec s \  (g_0\sigma_0+\frac{u}{2}(\sigma_0-\sigma_{z}))\vec S_1\cdot\vec s \  | \uparrow\rangle e^{-i\vec k\cdot\vec R}    \nonumber\\
+&   \left.  \sum\limits_{ k=0}^{k_{F\downarrow}}   
 \langle\downarrow| \vec S_2\cdot\vec s (g_0\sigma_0+\frac{d}{2}(\sigma_0+\sigma_{z}))\vec S_1\cdot\vec s \ | \downarrow\rangle e^{-i\vec k\cdot\vec R}  \right) + c.c.
 \label{ER3}
\end{align}  
This can be evaluated using the spin identities
\begin{align}
& {\rm Tr }   \big [\vec S_1\cdot \vec \sigma \  \vec S_2\cdot\vec \sigma\big ]=2\vec S_1\cdot\vec S_2,   \nonumber\\
& \langle\uparrow|    \vec S_1\cdot\vec \sigma   \   \vec S_2\cdot\vec \sigma \ | \uparrow\rangle=\vec S_1\cdot\vec S_2+i(\vec S_1\times\vec S_2)_{z},
\nonumber   \\
& \langle\downarrow|    \vec S_1\cdot\vec \sigma  \   \vec S_2\cdot\vec \sigma \ | \downarrow\rangle=\vec S_1\cdot\vec S_2-i(\vec S_1\times\vec S_2)_z, \nonumber\\
&\langle\uparrow|    \vec S_1\cdot\vec \sigma   \   \sigma_{z} \ \vec S_2\cdot\vec \sigma \ | \uparrow\rangle=-\vec S_1\cdot\vec S_2-i(\vec S_1\times\vec S_2)_{z}+2S_{1z}S_{2z}, \nonumber\\
&\langle\downarrow|   \vec S_1\cdot\vec \sigma \ \sigma_{z} \ \vec S_2\cdot\vec \sigma \  | \downarrow\rangle=\vec S_1\cdot\vec S_2-i(\vec S_1\times\vec S_2)_{z} - 2S_{1z}S_{2z},
\label{identities}
\end{align}
and the energy expression Eq. (\ref{ER3}) can be cast into the desired form 
\begin{equation}
E(R)=J \ \vec S_1\cdot \     \vec S_2+\vec D \  \cdot \  (\vec S_1\times\vec S_2)+\vec S_{1} \  \cdot  \ 
\stackrel{\leftrightarrow}{\Gamma}  \ \cdot \  \vec S_{2},
\label{ER4}
\end{equation}
where the various coefficients may be obtained 
 by converting the summations into integrations  in Eq. (\ref{ER3}).


\section{Results}
\subsection {Electron gas in 3D}

We now evaluate the interaction energy coefficients in Eq. (\ref{ER4}) for the spin polarized electron gas in 3D by evaluating the Green's function and performing the momentum integrations.
A straightforward contour integration\cite{Grosso} of the expression (\ref{Grr}) yields the Green's function
\begin{equation}
 G(\vec r, 0,E)
 =\frac{-2m}{4\pi r\hbar^{2}}
\begin{pmatrix}
e^{i\alpha(E+\Delta)r} & 0 \\
0 & e^{i\alpha(E-\Delta)r} 
 \end{pmatrix},
 \label{G-R0E}
\end{equation}
where
$  \alpha(x) =  (2m\hbar^{-2} x)^{1/2}$ for $ x>0$   and   
$  i (2m\hbar^{-2} |x|)^{1/2} $ for $ x < 0$.
Comparing Eq. (\ref{G-R0E}) with Eq. (\ref{cases}), we have:
$g_{0}= \beta e^{ikR}$, $ u = \beta e^{-\sqrt{2\delta^2-k^2}R}- g_0$, if $-\Delta<\varepsilon_{\vec k\uparrow}<\Delta$, or
$\beta e^{i\sqrt{k^2-2\delta^2}R}-g_0$, if $\varepsilon_{\vec k\uparrow}>\Delta$, and
$d=\beta  e^{i\sqrt{k^2+2\delta^2}R}- g_0$, if $\varepsilon_{\vec k\downarrow}>\Delta$. Here 
$\beta = -m (2 \pi \hbar^2 r )^{-1}$.
Performing the necessary integrations, we finally get from Eqs. (\ref{ER3}) and (\ref{identities}) the result for the RKKY interaction
\begin{equation}
J =  -  \frac{\lambda^2 m} {(4\pi)^3 R^4} \times (   I(x_\uparrow)+   I(x_\downarrow)),
\label{E16}
\end{equation}
where $ x_\uparrow = k_{F\uparrow} R$, $ x_\downarrow = k_{F\downarrow} R$, and  $I(x)= \sin(2x)-2x \cos(2x) $. Interestingly, the vector DM interaction $\vec D = 0$, while the tensor interaction survives:
\begin{equation}
\stackrel{\leftrightarrow}{\Gamma} =
\begin{pmatrix}
P & 0 & 0\\
0 & P & 0\\
0 & 0 & 0
 \end{pmatrix},
 \label{E17}
\end{equation}
where $
P=g(x_\downarrow, x_\uparrow) -J (x_\downarrow, x_\uparrow)   $ and defining 
$\eta=   (x_\uparrow^2-x_\downarrow^2)^{1/2} = \sqrt 2 \delta R$,  
the expression for $g$ is
\begin{align}
g(x_\downarrow, x_\uparrow)   & =   -\frac{\lambda^2 m}{8\pi^3 R^4} 
 \left(  \int_0^\eta   
e^{-\sqrt{\eta^2-x^2}}   + \int_{\eta}^{x_\uparrow}      \cos \sqrt{x^2-\eta^2}      \right.       \nonumber\\
 & \left.  + \int_{0}^{x_\downarrow}     \cos \sqrt{x^2+\eta^2}  \right) \times  x \sin x \  dx. 
 \label{E18}
\end{align}

We note that for the unpolarized electron gas, putting $k_{F\uparrow}=k_{F\downarrow}=k_{F}$, the above expressions yield  $P = 0$, so that we recover the standard RKKY result: 
$E(R)= J \ \vec S_1\cdot\vec S_2$, where 
$
J =   - \lambda^2 m/ ( 32\pi^3 R^4)    I(k_{F}R). 
$
We illustrate the oscillatory behavior of $J$ and $P$ in Fig. (\ref{RKKY}) for typical parameters.  
The beating pattern seen in the figure comes from the interference of the two Fermi momenta for the two spins. For example, it is easy to see from Eq. (\ref{E16}) that for large distances $J \sim   R^{-3} \cos ( k_{F\uparrow} R+ k_{F\downarrow}R) \times \cos ( k_{F\uparrow} R - k_{F\downarrow}R) $, which leads to the beating pattern as indicated in Fig. (\ref{RKKY}).

Another point is that the constant energy shift $E_0$, Eq. (\ref{E0-3D}),  is proportional to $\lambda \delta k_F$, where $\delta k_F  = k_{F\uparrow}  - k_{F\downarrow}$ is the spin polarization at the Fermi surface, while the magnetic interactions, Eqs. (\ref{E16} - \ref{E18}),  are proportional to the second order $\lambda^2$. Thus, the latter terms dominate for the weakly spin-polarized case, viz., $\delta k_F << (m k_F^2) \lambda$, where $k_F = (k_{F\uparrow}  + k_{F\downarrow})/2$ is the spin-averaged Fermi momentum and the typical scenario, $\delta k_F << k_F$, has been assumed. 
Finally, note from the form of the tensor interaction  Eq. (\ref{E17})  that the net interaction Eq. (\ref{ER4}) can be written as a sum of the Heisenberg and Ising interactions,
$E(R)=J^\prime  \ \vec S_1\cdot \     \vec S_2   +    J^{\prime \prime} S_{1z} S_{2z} $,
where $J^\prime = J +P$ and $J^{\prime \prime} = -P$. This is true both for the 3D electron gas as well as for the 2D gas described below.

\begin{figure}
\includegraphics[angle=0,width=0.48    \linewidth]{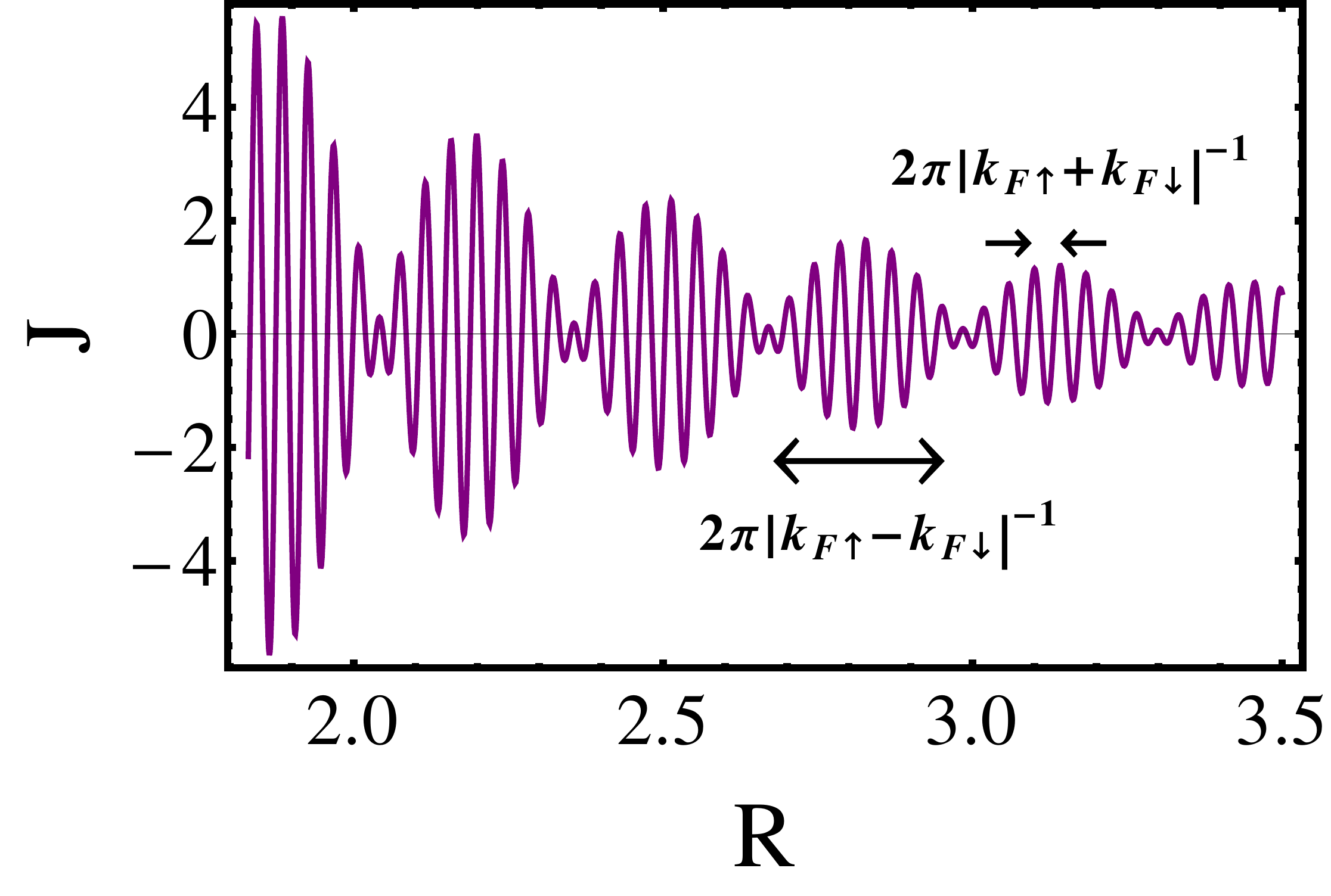}
\hfill
\includegraphics[angle=0,width=0.483    \linewidth]{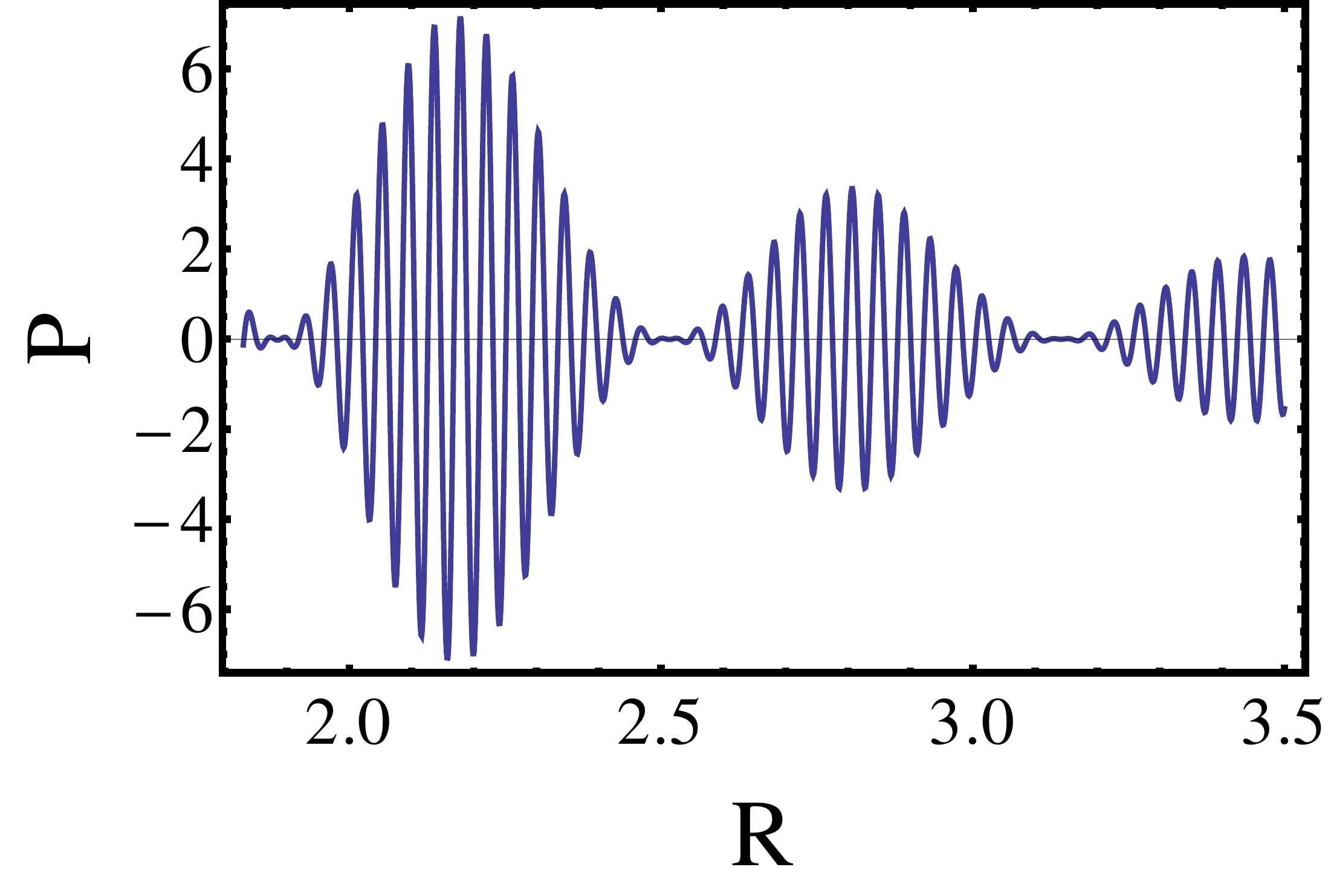}
\caption{(Color online)  
The RKKY interaction $J$ ({\it top}) and the tensor Dzyaloshinsky-Moriya interaction $P$ ({\it bottom}), in units of    
$m\pi^{-3}\lambda^{2}$, for the 3D
electron gas as a function of distance $R$, obtained from Eqs. (\ref{E16} - \ref{E18}). 
}
\label{RKKY}
\end{figure}

\subsection{Electron Gas in 2D}

In 2D,  the Green's function Eq. (\ref{Grr}) becomes
\begin{align}
 G_{\sigma\sigma^\prime}(\vec r,\vec r \ ^\prime,E)   
=\frac{\delta_{\sigma\sigma^\prime}}      {(2 \pi)^2}   
\int 
\frac{e^{i\vec k\cdot(\vec r-\vec r \  ^\prime)}}   {E+i\mu-\varepsilon_{\vec k ,\sigma}}  \ d^2k.
\end{align}
Using the Jacobi-Anger expansion\cite{Jacobi-Anger} in terms of the Bessel's functions, viz.,
$ e^{i \vec k \cdot \vec r}   =  J_0 (kr) + 2 \sum\limits_{n=1}^\infty i^n J_n (kr) \cos (n \theta)$, where $\theta$ is measured with respect to $\hat r$,
the angular integration produces just one non-zero term involving $J_0 (kr)$, and the radial integration finally leads to the well-known result\cite{Litvinov}:
\begin{equation}
G_{\sigma \sigma } (\vec r,\vec r \ ^\prime,E)  = -\frac{m}{\pi \hbar^2}K_0 \left[ -i  \frac{\sqrt {2m} }{\hbar}   |\vec r - \vec r \ ^\prime| (E \pm \Delta)^{1/2}   \right],
\label{2DGF}
\end{equation}
where + (-) is for $\sigma = \uparrow$ ($\downarrow$) and $K_0$ is the modified Bessel function of the second kind. 
For the perturbation of the $\vec k \sigma$ state, as before, we put $E = \varepsilon_{ \vec k \sigma}$ and express Eq. (\ref{2DGF}) as a function of the momentum $k$. The result is:
\begin{equation}
 G(\vec R, 0,\varepsilon_{ \vec k \uparrow})
 =\beta^\prime
\begin{pmatrix}
K_0 (-i k R)   & 0 \\
0 & K_0 (-i (k^2 - 2 \delta^2)^{1/2} R) 
 \end{pmatrix},
\end{equation}
where  $\delta = (2 m \Delta)^{1/2} \hbar^{-1}$ and $\beta^\prime = -m / (\pi \hbar^2 )$.
The same result for the down spin is:
\begin{equation}
 G(\vec R, 0,\varepsilon_{ \vec k \downarrow})
 = \beta^\prime
\begin{pmatrix}
K_0 (-i (k^2 + 2 \delta^2)^{1/2} R)   & 0 \\
0 &  K_0 (-i k R)
 \end{pmatrix}.
\end{equation}

	Writing this in terms of the Pauli matrices as in Eq. (\ref{cases}), we get: 
$g_{0}= \beta^\prime  K_0  (-ikR) $, 
$ u = \beta^\prime    K_0 $ $ (    \sqrt{2\delta^2 - k^2}    R) - g_0$, if $-\Delta  <\varepsilon_{\vec k\uparrow}   <\Delta$, and
$\beta^\prime    K_0 $ $ (-i  \sqrt {k^2 - 2\delta^2 }   R) - g_0 $, if $\varepsilon_{\vec k\uparrow}>\Delta$, and
$ d = \beta^\prime    K_0  $ $ (-i \sqrt { k^2 +  2\delta^2 } R) - g_0 $, 
if $\varepsilon_{\vec k\downarrow}>\Delta$. 
The remaining algebra is the same as the 3D case and the final results for the 2D electron gas reads:
\begin{align}
E_0^{2D} & = \frac{-\hbar\lambda}{8 \pi R^2 }(S_{1z}+ S_{2z})(x_\uparrow^2-x_\downarrow^2),  \\
J^{2D} & =  \frac{m\lambda^2 }  {16 \pi R^2} \times (I_2(x_\downarrow) + I_2(x_\uparrow) ),  \\
\vec D^{2D}  & = 0,
\label{E24}
\end{align}

%
%
%
%
%
%
\begin{align}
P^{2D}      & =    \frac{m\lambda^2 }  {8 \pi R^2} \times
 \left(       -   \frac{2}{\pi}    \int_0^\eta   
Re [ K_0 ( \sqrt{ \eta^2 - x^2}  ) ]
 + \int_{\eta}^{x_\uparrow}        Y_0 ( \sqrt{ x^2 - \eta^2 }  )      \right.
\notag\\ %
& \left. + \int_{0}^{x_\downarrow}        Y_0 ( \sqrt{ x^2 + \eta^2 }  )       \right) \times  x J_0 (x) \  dx - J^{2D} . 
\end{align}
%
where, again, $\eta^2 = 2 \delta^2 R^2$, and
$I_2 (x) = x^2 [J_0 (x) Y_0 (x) + J_1 (x) Y_1 (x)]$, the $J$'s and $Y$'s here being  the Bessel's and Neumann's functions, respectively.
In the large distance limit, $I_2 (x) \rightarrow -\pi^{-1} \sin (2x)$, so that the magnetic interactions fall off as $\sim R^{-2}$, with a beating pattern similar to the 3D case, as might be seen from Eq. (\ref{E24}) for $J^{2D}$. Once again, in the non-spin-polarized limit, $k_{F\uparrow} = k_{F\downarrow}$, we recover the standard results for the RKKY interaction for the 2D electron gas.

\section{Summary}

In summary, we extended the original RKKY description for the magnetic interaction in the electron gas to the spin-polarized case. The broken time-reversal symmetry produces, in addition to the RKKY interactions, the DM interaction terms, explicit expressions for which were derived for the cases of 2D and the 3D electron gas. Numerical results showed a beating pattern due to the interference between the two Fermi momenta $k_{F\uparrow}$ and  $k_{F\downarrow}$, for the two different spin channels. In the non-spin-polarized limit, $k_{F\uparrow} = k_{F\downarrow}$, the DM interaction vanishes and the standard result for the RKKY interaction $J$ for the electron gas is recovered.
Since for the spin-polarized electron gas, treated here, the inversion symmetry is not broken, the vector DM interaction term vanishes.
We note that in opposite situations where the inversion symmetry is broken, e.g., for the electron gas with Rashba or Dresselhaus terms, the  vector DM interaction
$\vec D$ is non-zero.\cite{Bruno,Chesi}

\section*{Acknowledgements}

We thank Mohammad Sherafati for helpful discussions and the U.S. Department of Energy, Office of Basic Energy Sciences, Division of Materials Sciences and Engineering for financial support under Award    No.DE-FG02-00ER45818.

\end{document}